\documentclass[seceq]{ptptex}
\usepackage{wrapft}
\usepackage{graphicx}

\newcommand{\ket}[1]{| {#1} \rangle}     
\newcommand{\kket}[1]{| {#1} \rangle\!\rangle}     
\newcommand{\rdket}[1]{|\!| {#1} )}     
\newcommand{\rrket}[1]{| {#1} )\!)}     
\newcommand{\dket}[1]{|\!| {#1} \rangle}     
\newcommand{\dkket}[1]{|\!| {#1} \rangle\!\rangle}     
\newcommand{\wtilde}[1]{\widetilde{#1}} 

\def\beq{\begin{eqnarray}}
\def\eeq{\end{eqnarray}}
\def\bsub{\begin{subequations}}
\def\esub{\end{subequations}}
\def\b{\begin{equation}}

\title{
Note on Many-Quark Model with $su(4)$ Algebraic Structure
}

\author{
Yasuhiko {\sc Tsue},$^{1}$ 
Constan\c{c}a {\sc Provid\^encia},$^{2}$ 
Jo\~ao da {\sc Provid\^encia}$^{2}$ and 
Masatoshi {\sc Yamamura}$^{3}$  
}

\inst{
$^{1}$Physics Division, Faculty of Science, Kochi University, Kochi 780-8520, 
Japan\\
$^{2}$Departamento de F\'{i}sica, Universidade de Coimbra, 3004-516 Coimbra, 
Portugal\\
$^{3}$Department of Pure and Applied Physics, 
Faculty of Engineering Science,\\
Kansai University, Suita 564-8680, Japan
\\

}

\recdate{
\today
}

\abst{
In a many-quark model developed in our previous paper where two-body color 
pairing and particle-hole type interactions are active, 
the exact energy eigenstates are re-formed with physically clearer 
expressions than those derived in our previous paper. 
By using the re-formed energy eigenstates, two types of the eigenstates 
in which the pairing correlation and the quark triplet formation 
separately appear definitely, are unified and this model can be treated 
for both the strong color correlations and the quark triplet formations. 
}

\begin{document}

\maketitle

\section{Introduction}

One of the recent interests of quark and hadron physics may be the 
investigation of the various properties which many-quark system 
reveals. 
In our previous paper\cite{1} which is referred to as (I), 
we have investigated the many-quark model where the two-body 
pairing interaction is active. 
This model is known as the Bonn model\cite{2} and it is known 
for this model to 
lead to the formation of the quark triplet 
while only two-body interaction is 
contained.\cite{3} 
As was pointed out in (I), the Bonn model has the $su(4)$ algebraic 
structure. 
Thus, from the viewpoint of the color $su(3)$-symmetry in 
quantum chromodynamics (QCD), this model can be extended to the 
many-quark model which does not have the $su(4)$ symmetry but has only the 
color $su(3)$-symmetry. 
In (I), we have formulated the above-mentioned idea in terms of the 
Schwinger boson representation for many-fermion system. 
As a result, we have given the many-quark model with the color pairing plus 
particle-hole type interactions under the color $su(3)$-symmetry, 
which we have called the modified Bonn model.

In (I), we gave the orthogonal set for the quantum states of this model 
in the boson realization for the many-quark system. 
Then, the exact eigenstates and eigenvalues for this modified Bonn model 
Hamiltonian were derived. 
As for the set of the eigenstates and the eigenenergies, there are two 
posibilities: 
One is classified as a form with the pairing correlation and the other 
is as a quark triplet formation.

In this paper, we reformulate the orthogonal set or the eigenstates 
for the modified Bonn model. 
Because the model is formulated in the boson space using the boson 
realization, the physical quantities peculiar to many-quark model 
are transcribed in boson representation. 
After that, the orthogonal set or the eigenstates are constructed in 
terms of the quark triplet, quark pair and single quark creation operators 
which are denoted as ${\hat B}^*$, ${\hat S}^i$ and ${\hat q}^i$ respectively, 
by acting on the minimum weight state for the $su(1,1)$ algebra appearing 
in the boson realization. 
By using the re-formed orthogonal set, which we denote as 
$\{\dket{lsr\omega}\}$, of the quantum states, 
it is shown that the eigenstates with the pairing-type correlation and 
with the quark triplet formation can be described in a unified way. 
This is one of our main purposes of this paper. 
Further, it is indicated that two types of the minimum weight states 
for the $su(1,1)$-algebra are necessary in order to construct the 
orthogonal set $\{\dket{lsr\omega}\}$. 
This fact may be an interesting feature for the model with the algebraic 
structure including the non-compact $su(1,1)$ sub-algebra.

As for the complete orthogonal set for the modified Bonn model, 
eight quantum numbers are needed. 
The complete orthogonal states are constructed from $\{\dket{lsr\omega}\}$, 
which degenerate with respect to the energy eigenvalues in the modified 
Bonn model. 
An idea to dissociate the degeneracy will be given and a possible method 
will be formulated by using the technique developed in the theory 
of the nuclear collective rotational motion.

This paper is organized as follows. 
In the next section, the outline of our modified Bonn model in the 
Schwinger boson representation is given by paying an attention to 
the algebraic structure of this many-quark system. 
In \S 3, three kinds of the basic building blocks to construct the 
orthogonal set are presented together with their physical and 
mathematical properties. 
The re-formed construction of the orthogonal set, 
$\{\dket{lsr\omega}\}$, is given in \S 4 by using the three kinds of the 
building blocks, namely, the quark triplet, the color pair and 
the single quark creations. 
In \S 5, the relevance to the two forms developed in (I) are indicated 
and these two forms are described in the unified way by means of the 
re-formed orthogonal set $\{\dket{lsr\omega}\}$. 
The expressions in terms of the degeneracy $\Omega$ and the quark 
numbers with color $i$, $N_i$, are given and the physical meaning 
of the state $\{\dket{lsr\omega}\}$ is also clarified in \S 6. 
The last section is devoted to the concluding remarks and 
the possible idea to dissociate the degeneracy with respect to the energy 
is mentioned in this last section. 
The proofs of various formulae and eigenvalue problem are given 
in Appendices A and B.

\section{Outline of the model in the 
Schwinger boson representation}

In this section, we will recapitulate the basic framework of many-quark model 
presented in (I). 
As was mentioned in \S 1, this model obeys the $su(4)$-algebra 
and is called the Bonn model. 
The Schwinger boson representation for the $su(4)$-algebra adopted in (I) 
is composed of eight kinds of bosons 
$({\hat a}, {\hat a}^*, {\hat b}, {\hat b}^*, 
{\hat a}_i, {\hat a}_i^*, {\hat b}_i, {\hat b}_i^*\ ;\ 
i=1,2,3)$, in which 
the symbol $i$ denotes the color $i$. 
With the use of the above bosons, the $su(4)$-generators are expressed 
in the form 
\beq\label{2-1}
& &{\hat S}^i={\hat a}_i^*{\hat b}-{\hat a}^*{\hat b}_i \ , \qquad
{\hat S}_i={\hat b}^*{\hat a}_i -{\hat b}_i^*{\hat a}  \ , \nonumber\\
& &{\hat S}_i^j=({\hat a}_i^*{\hat a}_j-{\hat b}_j^*{\hat b}_i)
+\delta_{ij}({\hat a}^*{\hat a}-{\hat b}^*{\hat b}) \ .
\eeq
Naturally, the Bonn model belongs to many-fermion models. 
We intend to describe this fermion system in the space spanned by 
the above boson operators. 
Therefore, it is indispensable to transcribe physical quantities 
peculiar to many-fermion model under investigation in boson space. 
Typical examples in the present case are ${\hat S}^1$, ${\hat S}^2$ 
and ${\hat S}^3$, which correspond to the quark-pair creations in 
colors (2, 3), (3, 1) and (1, 2), respectively. 
The degeneracy of the single-particle level $i$, $2\Omega$, is a 
parameter in the fermion space and it is positive integer. 
In the boson space, we treat this parameter as an operator 
$2{\hat \Omega}$, which is expressed in terms of the bosons. 
The quark number operator in the color $i$, ${\wtilde N}_i$, is also the same. 
The operator ${\wtilde N}_i$ should be transcribed in the boson space. 
We denote it as ${\hat N}_i$. 
In (I), we obtained ${\hat \Omega}$, ${\hat N}_i$ and ${\hat N}$ 
($=\sum_i{\hat N}_i$) in the following form: 
\beq
& &{\hat \Omega}=n_0+\frac{1}{2}\left[
({\hat a}^*{\hat a}+{\hat b}^*{\hat b})+
\sum_i({\hat a}_i^*{\hat a}_i+{\hat b}_i^*{\hat b}_i)
\right] \ , 
\label{2-2}\\
& &{\hat N}_i=n_0+{\hat a}^*{\hat a}+\sum_j{\hat a}_j^*{\hat a}_j
-({\hat a}_i^*{\hat a}_i-{\hat b}_i^*{\hat b}_i) \ , 
\label{2-3}\\
& &{\hat N}=3n_0+3{\hat a}^*{\hat a}+2\sum_i{\hat a}_i^*{\hat a}_i
+\sum_i{\hat b}_i^*{\hat b}_i \ . 
\label{2-4}
\eeq
Here, $n_0$ denotes a 
positive integer which corresponds to the seniority number 
in the $su(2)$-pairing model.

In the present boson space, there exists the $su(1,1)$-algebra which does not 
exist in the original fermion system: 
\b\label{2-5}
{\hat T}_{\pm,0}={\hat t}_{\pm,0}+{\hat \tau}_{\pm,0} \ .
\end{equation}
Here, ${\hat t}_{\pm,0}$ and ${\hat \tau}_{\pm,0}$ are defined as 
\beq
& &{\hat t}_+={\hat b}^*{\hat a}^* \ , \quad
{\hat t}_-={\hat a}{\hat b} \ , \quad
{\hat t}_0=\frac{1}{2}({\hat b}^*{\hat b}+{\hat a}^*{\hat a})+\frac{1}{2} \ , 
\label{2-6}\\
& &{\hat \tau}_+=\sum_i{\hat b}_i^*{\hat a}_i^* \ , \quad
{\hat \tau}_-=\sum_i{\hat a}_i{\hat b}_i \ , \quad
{\hat \tau}_0=\frac{1}{2}
\sum_i({\hat b}_i^*{\hat b}_i+{\hat a}_i^*{\hat a}_i)+\frac{3}{2} \ . 
\label{2-7}
\eeq
The sets $\{{\hat t}_{\pm,0}\}$ and $\{{\hat \tau}_{\pm,0}\}$ form two 
independent $su(1,1)$-algebras. 
The most important property of $\{{\hat T}_{\pm,0}\}$ is that 
$\{{\hat T}_{\pm,0}\}$ commutes with $\{{\hat S}^i,{\hat S}_i,{\hat S}_i^j\}$. 

We mentioned in (I) that as a sub-algebra, the present $su(4)$-algebra 
contains the $su(3)$-algebra, which includes the $su(2)$-algebra as a 
sub-algebra. 
The generators are listed as 
\beq
& &\hbox{\rm the\ $su(3)$-algebra}\quad
\{\ {\hat S}_1^2,\ {\hat S}_2^1,\ {\hat S}_1^3,\ {\hat S}_3^1, 
{\hat S}_2^3,\ {\hat S}_3^2, {\hat Q}_0, {\hat R}_0\ \} \ , 
\label{2-8}\\
& &\hbox{\rm the\ $su(2)$-algebra}\quad
\{\ {\hat R}_+,\ {\hat R}_-,\ {\hat R}_0 \} \ . 
\label{2-9}
\eeq
Here, ${\hat Q}_0$ and ${\hat R}_0$ are functions of $({\hat S}_1^1, 
{\hat S}_2^2, {\hat S}_3^3)$ and there are several possibilities. 
Depending on ${\hat R}_0$, 
the operators ${\hat R}_{\pm}$ are determined. 
In (I), we adopted the following form: 
\b\label{2-10}
{\hat Q}_0={\hat S}_1^1-\frac{1}{2}({\hat S}_2^2+{\hat S}_3^3) \ , 
\qquad
{\hat R}_0=\frac{1}{2}({\hat S}_2^2-{\hat S}_3^3) \ . 
\end{equation}
Then, ${\hat R}_\pm$ can be chosen as 
\b\label{2-11}
{\hat R}_+={\hat S}_2^3\ , \qquad
{\hat R}_-={\hat S}_3^2 \ .
\end{equation}
The Casimir operators denoted ${\hat {\mib P}}^2$, ${\hat {\mib Q}}^2$ and 
${\hat {\mib R}}^2$ for the $su(4)$-, the $su(3)$- and 
the $su(2)$-algebra are expressed, respectively, as 
\beq
& &{\hat {\mib P}}^2=
2({\hat S}^1{\hat S}_1+{\hat S}^2{\hat S}_2+{\hat S}^3{\hat S}_3)
+2({\hat S}_1^2{\hat S}_2^1+{\hat S}_1^3{\hat S}_3^1
+{\hat R}_+{\hat R}_-) 
\nonumber\\
& &\qquad
+\frac{3}{4}{\hat P}_0({\hat P}_0-4)+\frac{2}{3}{\hat Q}_0({\hat Q}_0-3)
+2{\hat R}_0({\hat R}_0-1) \ , 
\label{2-12}\\
& &{\hat {\mib Q}}^2=
2({\hat S}_1^2{\hat S}_2^1+{\hat S}_1^3{\hat S}_3^1
+{\hat R}_+{\hat R}_-)+\frac{2}{3}{\hat Q}_0({\hat Q}_0-3)
+2{\hat R}_0({\hat R}_0-1) \ , 
\label{2-13}\\
& &{\hat {\mib R}}^2=
{\hat R}_+{\hat R}_- +
{\hat R}_0({\hat R}_0-1) \ . 
\label{2-14}
\eeq
Here, ${\hat P}_0$ is defined as 
\b\label{2-15}
{\hat P}_0=\frac{1}{3}({\hat S}_1^1+{\hat S}_2^2+{\hat S}_3^3) \ .
\end{equation}

In (I), we investigated the following Hamiltonian: 
\beq
& &{\hat H}=-({\hat S}^1{\hat S}_1+{\hat S}^2{\hat S}_2
+{\hat S}^3{\hat S}_3) \ , 
\label{2-16}\\
& &{\hat H}_m={\hat H}+\chi{\hat {\mib Q}}^2\ . \qquad (\chi\ :\ 
{\hbox{\rm a\ real\ parameter}}) 
\label{2-17}
\eeq
The Hamiltonian ${\hat H}$ is given in the original Bonn model 
and a possible modification, $\chi{\hat {\mib Q}}^2$ is added in (I). 
The most fundamental relation in this model is as follows: 
\b\label{2-18}
[\ {\hat S}_i^j\ , \ {\hat H}\ ]=0 \ , \qquad 
[\ {\hat S}_i^j\ , \ {\hat H}_m\ ]=0 \ . \qquad (i,\ j=1,\ 2,\ 3)
\end{equation}
Here, the relation $[ {\hat S}_i^j , {\hat {\mib Q}}^2]=0$, should 
be noted. 
With the use of the relations (\ref{2-12}) and (\ref{2-13}), ${\hat H}$ 
can be expressed as 
\b\label{2-19}
{\hat H}=-\frac{1}{2}\left({\hat {\mib P}}^2-{\hat {\mib Q}}^2-\frac{3}{4}
{\hat P}_0({\hat P}_0-4)\right) \ .
\end{equation}
In this paper, we will investigate ${\hat H}$ and ${\hat H}_m$ 
under an idea which is different from that in (I).

\section{Three kinds of basic building blocks and their properties}

First, we note that in the present boson space, there exists one more 
$su(4)$-algebra, the generators of which are expressed as 
\beq\label{3-1}
& &{\hat q}^i={\hat b}_i^*{\hat b}-{\hat a}^*{\hat a}_i \ , \quad
{\hat q}_i={\hat b}^*{\hat b}_i-{\hat a}_i^*{\hat a} \ , \nonumber\\
& &{\hat q}_i^j=({\hat b}_i^*{\hat b}_j-{\hat a}_j^*{\hat a}_i)
-\delta_{ij}({\hat b}^*{\hat b}-{\hat a}^*{\hat a}) \ .
\eeq
Any generator $\{{\hat q}^i,{\hat q}_i,{\hat q}_i^j\}$ commutes with any of 
$\{{\hat T}_{\pm,0}\}$, but some generators do not commute with 
$\{{\hat S}^i,{\hat S}_i,{\hat S}_i^j\}$. 
With the use of ${\hat S}^i$ and ${\hat q}^i$, we define the following 
operators: 
\b\label{3-2}
{\hat B}^*=\sum_i{\hat S}^i{\hat q}^i \ , \qquad
{\hat B}=\sum_i{\hat q}_i{\hat S}_i \ .
\end{equation}
The operators $({\hat B}^*,{\hat B})$ also commutes with 
any of $\{{\hat T}_{\pm,0}\}$. 
Main aim of this section is to investigate various properties of the 
operators ${\hat S}^i$, ${\hat q}^i$ and ${\hat B}^*$. 
Through this task, we can conclude that these operators 
can be regarded as building blocks for constructing a possible form 
of the orthogonal set for the boson space. 

We notice, first, the following relation 
\b\label{3-3}
[\ {\hat S}^i \ , \ {\hat S}^j\ ]
=[\ {\hat S}^i \ , \ {\hat q}^j\ ]
=[\ {\hat q}^i \ , \ {\hat q}^j\ ]
=0 \ . \quad (\hbox{\rm any}\ i\ {\rm and}\ j)
\end{equation}
Then, we have 
\b\label{3-4}
[\ {\hat S}^i \ , \ {\hat B}^*\ ]
=[\ {\hat q}^i \ , \ {\hat B}^*\ ]
=0 \ . 
\end{equation}
From the above relations, we learn that the ordering of ${\hat S}^i$, 
${\hat q}^i$ and ${\hat B}^*$ for their products is arbitrary. 
Next, we show the properties of ${\hat S}^i$, ${\hat q}^i$ and ${\hat B}^*$ 
related to the $su(3)$- and the $su(2)$-generators shown in the relations 
(\ref{2-8}) and (\ref{2-9}), respectively. 
We know the relation 
\beq\label{3-5}
& &[\ {\hat S}_i^j \ , \ {\hat S}^k\ ]
=\delta_{ij}{\hat S}^k+\delta_{jk}{\hat S}^i \ , \nonumber\\
& &[\ {\hat S}_i^j \ , \ {\hat q}^k\ ]
=\delta_{ij}{\hat q}^k-\delta_{ik}{\hat q}^j \ .
\eeq
The relations (\ref{3-2}) and (\ref{3-5}) give us 
\b\label{3-6}
[\ {\hat S}_i^j \ , \ {\hat B}^*\ ]=0\ . 
\end{equation}
Then, we have 
\b\label{3-7}
[\ {\hat Q}_0 \ , \ {\hat B}^*\ ]=0\ , \qquad 
[\ {\hat R}_{\pm,0} \ , \ {\hat B}^*\ ]=0 \ . 
\end{equation}
The relations (\ref{3-6}) and (\ref{3-7}) tell us that ${\hat B}^*$ 
is a color-neutral operator, and naturally, $su(2)$-scalar. 
The operators ${\hat q}^i$ and ${\hat S}^i$ do not possess such 
properties. 
However, we notice ${\hat q}^1$ and ${\hat S}^3$. 
In the case ${\hat q}^1$, we have 
\beq
& &[\ {\hat S}_2^1 \ , \ {\hat q}^1\ ]=
[\ {\hat S}_3^1 \ , \ {\hat q}^1\ ]=0 \ ,\quad
[\ {\hat Q}_0 \ , \ {\hat q}^1\ ]=-{\hat q}^1 \ . 
\label{3-8}\\
& &[\ {\hat R}_{\pm,0} \ , \ {\hat q}^1\ ]=0 \ . 
\label{3-9}
\eeq
The relation (\ref{3-9}) tells us that ${\hat q}^1$ is $su(2)$-scalar, 
but, not color-neutral. 
The relation (\ref{3-8}) will play an important role 
for constructing the orthogonal set, which will be shown in the next section. 
The operator ${\hat S}^3$ obeys the relations 
\beq
& &[\ {\hat S}_2^1 \ , \ {\hat S}^3\ ]=
[\ {\hat S}_3^1 \ , \ {\hat S}^3\ ]=0 \ , \quad
[\ {\hat Q}_0 \ , \ {\hat S}^3\ ]=-\frac{1}{2}{\hat S}^3 \ , 
\label{3-10}\\
& &[\ {\hat R}_- \ , \ {\hat S}^3\ ]=0 \ , \quad
[\ {\hat R}_0 \ , \ {\hat S}^3\ ]=-\frac{1}{2}{\hat S}^3 \ .
\label{3-11}
\eeq
The above indicate that ${\hat S}^3$ is not color-neutral and also, 
not $su(2)$-scalar, 
but the relations (\ref{3-10}) and (\ref{3-11}) 
also play an important role for constructing the orthogonal set. 
The commutation relation of ${\hat B}^*$, ${\hat q}^1$ and ${\hat S}^3$ 
for ${\hat P}_0$ is given as 
\b\label{3-12}
[\ {\hat P}_0 \ , \ {\hat B}^*\ ]=2{\hat B}^*\ , \quad
[\ {\hat P}_0 \ , \ {\hat q}^1\ ]=\frac{2}{3}{\hat q}^1\ , \quad
[\ {\hat P}_0 \ , \ {\hat S}^3\ ]=\frac{4}{3}{\hat S}^3\ .
\end{equation}
The above are the mathematical properties of ${\hat S}^i$, ${\hat q}^i$ 
and ${\hat B}^*$.

Next, we investigate physical properties in relation to 
${\hat \Omega}$ and ${\hat N}$ presented in the relations (\ref{2-2}) and 
(\ref{2-4}), respectively. 
It is very easy to show the following relations: 
\beq
& &[\ {\hat \Omega} \ , \ {\hat B}^*\ ]
=[\ {\hat \Omega} \ , \ {\hat S}^3\ ]
=[\ {\hat \Omega} \ , \ {\hat q}^1\ ]
=0 \ , 
\label{3-13}\\
& &[\ {\hat N} \ , \ {\hat B}^*\ ]=3{\hat B}^* \ , \quad
[\ {\hat N} \ , \ {\hat S}^3\ ]=2{\hat S}^3 \ , \quad
[\ {\hat N} \ , \ {\hat q}^1\ ]={\hat q}^1 \ . 
\label{3-14}
\eeq
The relation (\ref{3-13}) shows that the degeneracy of the single-particle 
level does not depend on ${\hat B}^*$, ${\hat S}^3$ and ${\hat q}^1$. 
The relation (\ref{3-14}) tells us that the operators ${\hat B}^*$, 
${\hat S}^3$ and ${\hat q}^1$ change the quark number 3, 2 and 1, 
respectively. 
Therefore, it may be permitted for ${\hat B}^*$, ${\hat S}^3$ and 
${\hat q}^1$ to play a role of the quark-triplet, the quark-pair and 
single-quark creation, respectively. 
Combining with the mathematical properties, we have the following image: 
The quark-triplet is color-neutral and $su(2)$-scalar. 
The quark-pair is colored. The single-quark 
is colored, but $su(2)$-scalar. 
As a final remark of this section, we will mention the role of 
the boson operator ${\hat b}^*$ itself. 
The operator ${\hat b}^*$ obeys 
\b\label{3-15}
[\ {\hat \Omega} \ , \ {\hat b}^*\ ]=\frac{1}{2}{\hat b}^* \ , \quad
[\ {\hat N} \ , \ {\hat b}^*\ ]=0 \ . 
\end{equation}
The relation (\ref{3-15}) tells us that the degeneracy $\Omega$ is 
determined by ${\hat b}^*$, but, it does not change the quark number.

\section{Construction of the orthogonal set}

On the basis of the above argument in \S 3, we search the 
orthogonal set. 
We introduce the following state: 
\b\label{4-1}
\dket{lsr\omega}=({\hat S}^3)^{2l}({\hat q}^1)^{2s}
({\hat B}^*)^{2r}({\hat b}^*)^{2\omega}\ket{0} \ . 
\end{equation}
In this paper, we omit numerical factor 
such as normalization constant for any state except some cases. 
The state (\ref{4-1}) is an eigenstate for ${\hat P}_0$, ${\hat Q}_0$, 
${\hat R}_0$ and ${\hat T}_0$: 
\beq
& &{\hat P}_0\dket{lsr\omega}
=-2\left(\omega-2\left(r+\frac{1}{3}(2l+s)\right)\right)
\dket{lsr\omega} \ , 
\label{4-2}\\
& &{\hat Q}_0\dket{lsr\omega}
=-(l+2s)\dket{lsr\omega} \ , \qquad
{\hat R}_0\dket{lsr\omega}
=-l\dket{lsr\omega} \ , 
\label{4-3}\\
& &{\hat T}_0\dket{lsr\omega}
=(\omega+2)\dket{lsr\omega} \ . 
\label{4-4}
\eeq
The above indicate that the set $\{\dket{lsr\omega}\}$ 
forms an orthogonal set. 
However, our present system is composed by the eight kinds of bosons, 
and then, the complete set is specified by eight quantum 
numbers. 
In order to obtain the complete orthogonal set, we pay attention to the 
relations 
\beq
& &{\hat S}_2^1\dket{lsr\omega}={\hat S}_3^1\dket{lsr\omega}=
{\hat R}_-\dket{lsr\omega}=0 \ , 
\label{4-5}\\
& &{\hat T}_-\dket{lsr\omega}=0 \ . 
\label{4-6}
\eeq
The relations (\ref{4-5}) and (\ref{4-6}) tell us that the state 
$\dket{lsr\omega}$ is the minimum weight state for the $su(3)$- and 
the $su(1,1)$-algebra. 
Then, a possible complete orthogonal set is given in the form 
\b\label{4-7}
\dket{k\kappa\kappa_0 lsr\omega\omega_0}
=({\hat T}_+)^{\omega_0-\omega}{\hat Q}_+(lk\kappa\kappa_0)\dket{lsr\omega} \ .
\end{equation}
Here, ${\hat Q}_+(lk\kappa\kappa_0)$ is a certain function of ${\hat S}_1^2$, 
${\hat S}_1^3$ and ${\hat R}_+$ and the concrete expression is given in the 
relation (I$\cdot$A$\cdot$8). 
Instead of the form (\ref{4-7}), we can adopt the following form: 
\beq\label{4-8}
& &\dket{k\kappa\kappa_0 lsr\omega\omega_0}
=({\hat T}_+)^{\omega_0-\omega}({\hat R}_+)^{\kappa+\kappa_0}
({\hat S}_1^2)^{k-l+\kappa}({\hat S}_1^4)^{k+l-\kappa}
\dket{lsr\omega}\ . 
\eeq
The operator ${\hat S}_1^4$ is defined as 
\b\label{4-9}
{\hat S}_1^4={\hat S}_1^3{\hat R}_0-\frac{1}{2}
{\hat S}_1^2{\hat R}_+ \ . \quad 
([\ {\hat S}_1^4\ , \ {\hat S}_1^2\ ]=0)
\end{equation}
In Appendix, we will give the reason why the expression (\ref{4-8}) 
is permitted. 
The state $\dket{k\kappa\kappa_0lsr\omega\omega_0}$ is an eigenstate of 
${\hat {\mib P}}^2$, ${\hat P}_0$, ${\hat {\mib Q}}^2$, ${\hat Q}_0$, 
${\hat {\mib R}}^2$, ${\hat R}_0$, ${\hat {\mib T}}^2$ and ${\hat T}_0$, 
the eigenvalues of which are summarized as follows: 
\bsub\label{4-10}
\beq
& &{\hat {\mib P}}^2\ :\ (2(r+s)-\omega)^2+2\omega(\omega+3) \ , \quad
{\hat P}_0 \ : \ 2\left(2\left(r+\frac{1}{3}(2l+s)\right)-\omega\right) \ , 
\nonumber\\
& &{\hat {\mib Q}}^2\ :\ \frac{2}{3}(l+2s)(l+2s+3)+2l(l+1) \ , \quad
{\hat Q}_0 \ : \ 3k-(l+2s) \ , 
\nonumber\\
& &{\hat {\mib R}}^2\ :\ \kappa(\kappa+1) \ , \quad
{\hat R}_0 \ : \ \kappa_0 \ , 
\label{4-10a}\\
& &{\hat {\mib T}}^2\ :\ (\omega+2)(\omega+1) \ , \quad
{\hat T}_0 \ : \ \omega+2 \ . 
\label{4-10b}
\eeq
\esub
Straightforward calculation gives us the above eigenvalues except 
the case ${\hat {\mib P}}^2$. 
In Appendix, we will show the derivation of the eigenvalue of 
${\hat {\mib P}}^2$. 
The quantum numbers obey the conditions 
\bsub\label{4-11}
\beq
& &l+s+r \leq \omega \ , 
\label{4-11a}\\
& &k\leq s\ , 
\label{4-11b}\\
& &|l-k| \leq \kappa \leq l+k \ , 
\label{4-11c}\\
& &-\kappa \leq \kappa_0 \leq \kappa\ , \qquad
\omega \leq \omega_0 \ . 
\label{4-11d}
\eeq
\esub
The conditions (\ref{4-11a}), (\ref{4-11b}) and (\ref{4-11c}) are discussed 
in Appendix. 
The condition (\ref{4-11d}) comes from the rules of the $su(2)$- and 
the $su(1,1)$-algebra. 
The conditions (\ref{4-11a}) and (\ref{4-11b}) will play a central role in 
\S\S 5 and 7, respectively.

The relation (\ref{2-18}) teaches us that the energy eigenvalue is determined 
by $\dket{lsr\omega}$. 
Therefore, we pay a special attention to this state. 
The state $\dket{lsr\omega}$ is constructed by operating the building blocks 
${\hat S}^3$, ${\hat q}^1$ and ${\hat B}^*$ for $2l$-, $2s$- and $2r$-times, 
respectively, on the state $({\hat b}^*)^{2\omega}\ket{0}$. 
Of course, the ordering of the operation is arbitrary and each building block 
plays its own role which was discussed in \S 3. 
However, $({\hat S}_3,{\hat S}^3)$, $({\hat q}_1,{\hat q}^1)$ and 
$({\hat B},{\hat B}^*)$ do not behave independently from one another. 
For example, we have 
\beq
& &[\ [\ {\hat q}_1 \ , \ {\hat S}^3\ ] \ , \ {\hat S}^3 \ ]=0 \ , \quad
{\hat q}_1({\hat b}^*)^{2\omega}\ket{0}
=[\ {\hat q}_1\ , \ {\hat S}^3\ ]({\hat b}^*)^{2\omega}\ket{0}=0 \ , 
\label{4-12}\\
& &[\ [\ {\hat S}_3 \ , \ {\hat q}^1\ ] \ , \ {\hat q}^1 \ ]=0 \ , \quad
{\hat S}_3({\hat b}^*)^{2\omega}\ket{0}
=[\ {\hat S}_3\ , \ {\hat q}^1\ ]({\hat b}^*)^{2\omega}\ket{0}=0 \ . 
\label{4-13}
\eeq
The relations (\ref{4-12}) and (\ref{4-13}) give us 
\b\label{4-14}
{\hat q}_1({\hat S}^3)^{2l}({\hat b}^*)^{2\omega}\ket{0}=0 \ , \qquad
{\hat S}_3({\hat q}^1)^{2s}({\hat b}^*)^{2\omega}\ket{0}=0 \ .
\end{equation}
Therefore, if restricting 
$\{({\hat S}^3)^{2l}({\hat q}^1)^{2s}({\hat b}^*)^{2\omega}\ket{0}\}$, 
$({\hat S}_3,{\hat S}^3)$ and $({\hat q}_1,{\hat q}^1)$ behave 
independently of each other. 
But, if $({\hat B},{\hat B}^*)$ is included, the situation becomes 
complicated. 
For example, we have the relation 
\beq\label{4-15}
& &{\hat q}_1({\hat B}^*)^{2r}({\hat b}^*)^{2\omega}\ket{0}
={\hat S}^1({\hat B}^*)^{2r-1}({\hat b}^*)^{2\omega}\ket{0} \ , \nonumber\\
& &{\hat S}_3({\hat B}^*)^{2r}({\hat b}^*)^{2\omega}\ket{0}
={\hat q}^3({\hat B}^*)^{2r-1}({\hat b}^*)^{2\omega}\ket{0} \ . 
\eeq
The above tells us that our building blocks are not elementary 
but composite. 
In relation to the above consideration, the work presented by Pittel 
et al. may be interesting.

Finally, we present the energy eigenvalue based on the use of the expression 
(\ref{2-19}). 
The state $\dket{lsr\omega}$ is the eigenstate of ${\hat {\mib P}}^2$ and 
${\hat {\mib Q}}^2$, the eigenvalues of which are the same as those shown 
in the relation (\ref{4-10a}). 
Further, $\dket{lsr\omega}$ is also the eigenstate of ${\hat P}_0$, the 
eigenvalue of which is shown in the relation (\ref{4-2}). 
Under the above consideration, the form (\ref{2-19}) gives us the following 
energy eigenvalue: 
\bsub\label{4-16}
\beq
& &E_{lsr\omega}^{(m)}=E_{lsr\omega}+\chi F_{ls} \ , 
\label{4-16a}\\
& &E_{lsr\omega}=-[2l(2(\omega-s-2r-l)+1)+2r(2(\omega-r)+3)] \ , 
\label{4-16b}\\
& &F_{ls}=2l(l+1)+\frac{2}{3}(l+2s)(l+2s+3) \ .
\label{4-16c}
\eeq
\esub
Concerning the energy eigenvalue (\ref{4-16a}), we must note the 
following: 
the operators $({\hat T}_+,{\hat Q}_+(lk\kappa\kappa_0))$ and 
$({\hat T}_+,{\hat R}_+,{\hat S}_1^2,{\hat S}_1^4)$ appearing in the 
forms (\ref{4-7}) and (\ref{4-8}), respectively, commute with 
${\hat H}$ and ${\hat H}_m$ in the relations (\ref{2-16}) and 
(\ref{2-17}), respectively. 
Therefore, the state $\dket{k\kappa\kappa_0 lsr\omega\omega_0}$ gives 
us the same energy eigenvalue as that given by $\dket{lsr\omega}$. 
This means that these states are degenerate. 
Of course, there exist exceptions. 
Subject on this degeneracy will be taken up again in \S\S 6 and 7.

\section{Relevance to the two forms presented in (I)}

As was mentioned in \S 1, in (I), we have investigated the present system 
under the two forms. 
In this section, we will examine the relevance of the state (\ref{4-1}) 
to the states (I$\cdot$3$\cdot$25) and (I$\cdot$4$\cdot$24). 
First, we discuss the case of the state (I$\cdot$4$\cdot$24), 
which is given as 
\b\label{5-1}
\ket{\lambda\rho\sigma_0\sigma_1}
=({\hat S}^3)^{2\lambda}({\hat S}^4)^{2\rho}
({\hat b}_1^*)^{2(\sigma_1-\sigma_0)}({\hat b}^*)^{2\sigma_0}\ket{0}\ . 
\end{equation} 
Here, we omit the interpretation of the quantum numbers. 
The operator ${\hat S}^4$ is defined as 
\b\label{5-2}
{\hat S}^4
={\hat S}^1{\hat Q}_0+{\hat S}^2{\hat S}_1^2+{\hat S}^3{\hat S}_1^3 
\ . 
\end{equation}
In (I), ${\hat S}^4$ played a central role for the description 
of the present system under the name of the pairing correlation. 
Our aim is to show that the state (\ref{5-1}) can be re-formed to the 
state (\ref{4-1}).

First, we note the following relation: 
\b\label{5-3}
[\ {\hat q}^1 \ ,\ {\hat S}^4\ ]={\hat B}^*\ , \quad
[\ {\hat B}^* \ ,\ {\hat S}^4\ ]=0\ , \quad
[\ {\hat B}^* \ ,\ {\hat q}^1\ ]=0\ .
\end{equation}
Further, we notice the formula
\b\label{5-4}
({\hat S}^4)^{n}({\hat q}^1)^{n}
=(-)^n({\hat B}^*-{\hat q}^1{\hat S}^4)(2{\hat B}^*-{\hat q}^1{\hat S}^4)
\cdots (n{\hat B}^*-{\hat q}^1{\hat S}^4) \ .
\end{equation}
Proof of the formula (\ref{5-4}) can be performed by the mathematical 
induction with the relation (\ref{5-3}) and the relation shown as 
\b\label{5-5}
(m{\hat B}^*-{\hat q}^1{\hat S}^4){\hat q}^1
={\hat q}^1((m+1){\hat B}^*-{\hat q}^1{\hat S}^4) \ .
\end{equation}
With the use of the formula (\ref{5-4}), for $m \geq 0$, we have 
\beq\label{5-6}
({\hat S}^4)^n({\hat q}^1)^{n+m}({\hat b}^*)^p\ket{0}
&=&({\hat S}^4)^n({\hat q}^1)^n\cdot
({\hat q}^1)^m({\hat b}^*)^p\ket{0} \nonumber\\
&=&(-)^n\frac{(n+m)!}{m!}({\hat B}^*)^n\cdot ({\hat q}^1)^m({\hat b}^*)^p
\ket{0} \ .
\eeq
Here, we used 
\beq
& &{\hat S}^4({\hat b}^*)^p\ket{0}=0 \ , 
\label{5-7}\\
& &(k{\hat B}^*-{\hat q}^1{\hat S}^4)({\hat q}^1)^m({\hat b}^*)^p\ket{0} 
=(k+m){\hat B}^*({\hat q}^1)^m({\hat b}^*)^p \ket{0} \ .
\label{5-8}
\eeq
Further, for $m\geq 1$, we have 
\beq\label{5-9}
& &({\hat S}^4)^{m+n}({\hat q}^1)^{n}({\hat b}^*)^p\ket{0}
=({\hat S}^4)^m\cdot ({\hat S}^4)^n({\hat q}^1)^n({\hat b}^*)^p\ket{0}
\nonumber\\
&=&({\hat S}^4)^m\cdot (-)^n({\hat B}^*)^n({\hat b}^*)^p\ket{0}
=(-)^n({\hat B}^*)^n({\hat S}^4)^m({\hat b}^*)^p\ket{0}
=0 \ .
\eeq
With the use of the operator ${\hat q}^1$, the state (\ref{5-1}) 
can be expressed as 
\b\label{5-10}
\ket{\lambda\rho\sigma_0\sigma_1}
=\frac{(2\sigma_0)!}{(2\sigma_1)!}({\hat S}^3)^{2\lambda}
({\hat S}^4)^{2\rho}({\hat q}^1)^{2(\sigma_1-\sigma_0)}
({\hat b}^*)^{2\sigma_1}\ket{0} \ .
\end{equation}
By reading $n=2\rho$, $n+m=2(\sigma_1-\sigma_0)$ and $p=2\sigma_1$ 
in the formula (\ref{5-6}), the state (\ref{5-10}) can be expressed in the 
form 
\beq\label{5-11}
\ket{\lambda\rho\sigma_0\sigma_1}&=&
({\hat S}^3)^{2\lambda}({\hat B}^*)^{2\rho}
({\hat q}^1)^{2(\sigma_1-\sigma_0)-2\rho}({\hat b}^*)^{2\sigma_1}\ket{0}
\nonumber\\
&=&({\hat S}^3)^{2\lambda}
({\hat q}^1)^{2(\sigma_1-\sigma_0-\rho)}({\hat B}^*)^{2\rho}
({\hat b}^*)^{2\sigma_1}\ket{0} \nonumber\\
&=&\dket{\lambda,\sigma_1-\sigma_0-\rho,\rho,\sigma_1} \ .
\eeq
Here, we used $[{\hat B}^*,{\hat q}^1]=0$ and omitted the numerical 
factor such as factorial. 
We can see that the quantum numbers $l$, $s$, $r$ and $\omega$ in the 
state (\ref{4-1}) is nothing but 
\b\label{5-12}
l=\lambda\ , \quad s=\sigma_1-\sigma_0-\rho \ , \quad
r=\rho \ , \quad \omega=\sigma_1 \ .
\end{equation}
Then, the energy eigenvalue (\ref{4-16}) becomes the same expression 
as that shown in the form (I$\cdot$4$\cdot$45) with (I$\cdot$4$\cdot$43b).

Next, we will investigate the relevance to the state (I$\cdot$3$\cdot$25), 
which is copied in the form
\beq
& &\dkket{\lambda\tau tT}=\left({\hat O}_+(t\tau)\right)^{T-(t+\tau)}
\dket{\lambda\tau}\otimes \kket{t} \ , 
\label{5-13}\\
& &\dket{\lambda\tau}=({\hat a}_3^*)^{2\lambda}
({\hat b}_1^*)^{2\tau-3-2\lambda}\ket{0} \ , \qquad
\kket{t}=({\hat b}^*)^{2t-1}\ket{0}\ . 
\label{5-14}
\eeq
Here, ${\hat O}_+(t\tau)$ is defined in the relation (I$\cdot$3$\cdot$22). 
The states $\dkket{\lambda\tau tT}$, $\dket{\lambda\tau}$ and $\kket{t}$ 
obey 
\beq
& &{\hat T}_-\dkket{\lambda\tau tT}=0 \ , \qquad
{\hat T}_0\dkket{\lambda\tau tT}=T\dkket{\lambda\tau tT} \ , 
\label{5-15}\\
& &{\hat \tau}_-\dket{\lambda\tau}=0 \ , \qquad
{\hat \tau}_0\dket{\lambda\tau}=\tau\dket{\lambda\tau} \ , 
\label{5-16}\\
& &{\hat t}_-\kket{t}=0 \ , \qquad
{\hat t}_0\kket{t}=t\kket{t} \ . 
\label{5-17}\\
& &\qquad
T=t+\tau, \ t+\tau+1,\ t+\tau+2, \cdots .
\label{5-18}
\eeq
The states $\dkket{\lambda\tau tT}$, $\dket{\lambda\tau}$ and $\kket{t}$ 
are the minimum weight states of the $su(1,1)$-algebra, 
$\{{\hat T}_{\pm,0}\ (={\hat \tau}_{\pm,0}+{\hat t}_{\pm,0})\}$, 
$\{{\hat \tau}_{\pm,0}\}$ and $\{{\hat t}_{\pm,0}\}$ specified by 
$T_0=T$, $\tau_0=\tau$ and $t_0=t$, respectively. 
The $(T-(t+\tau))$-times operation of ${\hat O}_+(t\tau)$ on the state 
$\dket{\lambda\tau}\otimes \kket{t}$ gives us the state 
$\dkket{\lambda\tau tT}$.

On the other hand, the state (\ref{4-1}) can be rewritten as 
\beq\label{5-19}
\dket{lsr\omega}&=&
({\hat B}^*)^{2r}({\hat S}^3)^{2l}({\hat q}^1)^{2s}({\hat b}^*)^{2\omega}
\ket{0} \nonumber\\
&=&({\hat B}^*)^{2r}({\hat a}_3^*)^{2l}({\hat b}_1^*)^{2s}
({\hat b}^*)^{2(\omega-(s+l))}\ket{0} \ .
\eeq
The state $\dket{lsr\omega}$ satisfies
\b\label{5-20}
{\hat T}_-\dket{lsr\omega}=0 \ , \quad
{\hat T}_0\dket{lsr\omega}=T\dket{lsr\omega}\ , \quad
T=\omega+2 \ .
\end{equation}
Further, $\dket{lsr\omega}$ can be rewritten in the form 
\b\label{5-21}
\dket{lsr\omega}=({\hat B}^*)^{2r}({\hat b}^*)^{4r}\dket{l\tau}\otimes 
\kket{t}\ . 
\end{equation}
Here, $\tau$ and $t$ are defined through 
\b\label{5-22}
2\tau-3=2(s+l)\ , \qquad 2t-1=2(\omega-(l+s)-2r) \ .
\end{equation}
The relations (\ref{5-20}) and (\ref{5-22}) give us 
\b\label{5-23}
T=t+\tau+2r \ , \quad {\rm i.e.}, \quad 2r=T-(t+\tau) \ .
\end{equation}
Since $\dket{lsr\omega}$ is the minimum weight state of the $su(1,1)$-algebra 
$\{{\hat T}_{\pm,0}\}$, we have the following expression: 
\b\label{5-24}
\dket{lsr\omega}=\left({\hat O}_+(t\tau)\right)^{T-(t+\tau)}
\dket{l\tau}\otimes \kket{t} \ .
\end{equation}
The state (\ref{5-24}) is the state (\ref{5-13}) itself. 
It may be interesting to see that the operation of 
$({\hat B}^*)^{2r}({\hat b}^*)^{4r}$ on the state 
$\dket{l\tau}\otimes \kket{t}$ is equivalent to that of 
$({\hat O}_+(t\tau))^{T-(t+\tau)}$.

From the above argument, we have the correspondence 
\b\label{5-25}
l=\lambda\ , \quad
s=\frac{1}{2}(2\tau-3-2\lambda) \ , \quad
r=\frac{1}{2}(T-t-\tau) \ , \quad
\omega=T-2\ .
\end{equation}
Of course, the energy eigenvalue (\ref{4-16}) under the relation 
(\ref{5-25}) is identical to the form (I$\cdot$3$\cdot$29). 
However, we must notice the re-formation from the form (\ref{5-19}) 
to the form (\ref{5-21}). 
Since $2t-1 \geq 0$, the relation (\ref{5-22}) gives us 
\b\label{5-26}
r \leq \frac{1}{2}(\omega-(l+s)) \ . 
\end{equation}
This means that the form (\ref{5-21}) is valid under the condition 
(\ref{5-26}). 
However, our present system obeys the condition (\ref{4-11a}), which 
is rewritten as $r \leq \omega-(l+s)$. 
Therefore, inevitably, we must investigate the case 
$(\omega-(l+s))/2 \leq r \leq \omega-(l+s)$.

In connection with the state $\dket{lsr\omega}$, we introduce the 
following state in the present boson space: 
\b\label{5-27}
\rdket{lsr\omega}=({\hat q}_3)^{2l}({\hat S}_1)^{2s}
({\hat B})^{2(\omega-(l+s)-r)}({\hat a}^*)^{2\omega}\ket{0} \ .
\end{equation}
The exponent $2(\omega-(l+s)-r)$ in the state (\ref{5-27}) 
should be positive or zero, and then, we have 
\b\label{5-28}
r \leq \omega-(l+s) \ .
\end{equation}
The relation (\ref{5-28}) is nothing but the condition (\ref{4-11a}). 
Straightforward calculation gives us the following relations: 
\beq
& &{\hat P}_0\rdket{lsr\omega}=-2\left[
\omega-2\left(r+\frac{1}{3}(2l+s)\right)\right]\rdket{lsr\omega} \ , 
\label{5-29}\\
& &{\hat Q}_0\rdket{lsr\omega}=-(l+2s)\rdket{lsr\omega} \ , \qquad
{\hat R}_0\rdket{lsr\omega}=-l\rdket{lsr\omega} \ , 
\label{5-30}\\
& &{\hat T}_0\rdket{lsr\omega}=(\omega+2)\rdket{lsr\omega}  \ , 
\label{5-31}\\
& &{\hat S}_2^1\rdket{lsr\omega}={\hat S}_3^1\rdket{lsr\omega}=
{\hat R}_-\rdket{lsr\omega}=0 \ , 
\label{5-32}\\
& &{\hat T}_-\rdket{lsr\omega}=0 \ . 
\label{5-33}
\eeq
If we compare the relations (\ref{5-29})$\sim$(\ref{5-33}) with the 
relations (\ref{4-2})$\sim$(\ref{4-6}), the state 
$\rdket{lsr\omega}$ should be identical with the state 
$\dket{lsr\omega}$ except the normalization constant: 
\b\label{5-34}
\rdket{lsr\omega}=\dket{lsr\omega} \ .
\end{equation}
Of course, we can define $\rdket{k\kappa\kappa_0 lsr\omega\omega_0}$ 
in the same form as $\dket{k\kappa\kappa_0 lsr\omega\omega_0}$. 
The state $\rdket{lsr\omega}$ is the re-formed expression of 
$\dket{lsr\omega}$.

Now, in the same idea as that for the form (\ref{5-19}), we can 
rewrite the state (\ref{5-27}): 
\b\label{5-35}
\rdket{lsr\omega}=({\hat B})^{2(\omega-(l+s)-r)}
({\hat a}^*)^{4(\omega-(l+s)-r)} \dket{l\tau}\otimes \rrket{t} \ .
\end{equation}
Here, $\dket{l\tau}$ and $\tau$ are given in the relations 
(\ref{5-14}) and (\ref{5-22}), respectively, and $\rrket{t}$ and $t$ 
are defined as 
\b\label{5-36}
\rrket{t}=({\hat a}^*)^{2t-1}\ket{0} \ , \qquad
2t-1=2(2r+(l+s)-\omega) \ . 
\end{equation}
Of course, as is shown in the relations (\ref{5-31}) and (\ref{5-33}), 
$\rdket{lsr\omega}$ is the minimum state for $\{{\hat T}_{\pm,0}\}$ with 
$T=\omega+2$. 
In the same idea as the relation (\ref{5-23}), we have 
\b\label{5-37}
T=t+\tau+2(\omega-(l+s)-r) \ , \quad
{\rm i.e.},\quad 
2(\omega-(l+s)-r)=T-(t+\tau) \ .
\end{equation}
Therefore, $\rdket{lsr\omega}$ can be expressed in the form 
\b\label{5-38}
\rdket{lsr\omega}=({\hat O}_+(t\tau))^{T-(t+\tau)}
\dket{l\tau}\otimes \rrket{t} \ .
\end{equation}
The conditions $2t-1 \geq 0$ for the relation (\ref{5-36}) and (\ref{5-28}) 
give us the following inequality: 
\b\label{5-39}
\frac{1}{2}(\omega-(l+s)) \leq r \leq \omega-(l+s) \ .
\end{equation} 
Thus, we arrived at our aim.

From the above argument, we have the correspondence 
\b\label{5-40}
l=\lambda\ , \quad 
s=\frac{1}{2}(2\tau-3-2\lambda) \ , \quad
r=\frac{1}{2}(T-(1-t)-\tau) \ , \quad
\omega=T-2 \ .
\end{equation}
The correspondence (\ref{5-40}) should be compared with that shown in the 
relation (\ref{5-25}). 
As was demonstrated in the relations (I$\cdot$3$\cdot$43)$\sim$
(I$\cdot$3$\cdot$47), the quantum number $t$ is replaced with $(1-t)$ 
in the quantum number $r$. 
Of course, the energy eigenvalue (\ref{4-16}) under the relation (\ref{5-40}) 
becomes to the form (I$\cdot$3$\cdot$48).

\section{Expressions in terms of the degeneracy $\Omega$ and the quark 
numbers $N_1$, $N_2$ and $N_3$}

As were shown in the relations (\ref{4-2})$\sim$(\ref{4-4}), 
the state $\dket{lsr\omega}$ is an eigenstate of ${\hat P}_0$, 
${\hat Q}_0$, ${\hat R}_0$ and ${\hat T}_0$. 
But, the eigenvalues of these operators are not directly connected to the 
present many-quark model. 
In order to make the connection, 
we take up the operators ${\hat \Omega}$, ${\hat N}_1$, ${\hat N}_2$ and 
${\hat N}_3$ presented in the relations (\ref{2-2}) and (\ref{2-3}). 
For obtaining their eigenvalues, the following formulae are useful: 
\beq
& &[\ {\hat \Omega}\ , \ {\hat S}^3\ ]=[\ {\hat \Omega}\ , \ {\hat q}^1\ ]
=[\ {\hat \Omega}\ , \ {\hat B}^*\ ]=0 \ , \nonumber\\
& &[\ {\hat \Omega}\ , \ {\hat b}^*\ ]=\frac{1}{2}{\hat b}^* \ , \quad
{\hat \Omega}\ket{0}=n_0\ket{0} \ , 
\label{6-1}\\
& &[\ {\hat N}_i\ , \ {\hat S}^3\ ]=(1-\delta_{i3}){\hat S}^3 \ , \quad
[\ {\hat N}_i\ , \ {\hat q}^1\ ]=\delta_{i1}{\hat q}^1 \ , \quad
[\ {\hat N}_i\ , \ {\hat B}^*\ ]={\hat B}^* \ , \nonumber\\
& &[\ {\hat N}_i\ , \ {\hat b}^*\ ]=0 \ , \quad
{\hat N}_i\ket{0}=n_0\ket{0} \ . 
\label{6-2}
\eeq
Here, some of the above relations are copied from the relations 
(\ref{3-13}) and (\ref{3-15}). 
The state $\dket{lsr\omega}$ is 
the eigenstate of 
${\hat \Omega}$, ${\hat N}_1$, ${\hat N}_2$ and ${\hat N}_3$ and 
their eigenvalues 
$\Omega$, $N_1$, $N_2$ and $N_3$ 
are given in the form 
\beq
& &\Omega=n_0+\omega \ , 
\label{6-3}\\
& &N_1=n_0+2l+2s+2r \ , \quad
N_2=n_0+2l+2r \ , \quad
N_3=n_0+2r \ . 
\label{6-4}
\eeq
Conversely, we have 
\beq
& &\omega=\Omega-n_0 \ , 
\label{6-5}\\
& &2s=N_1-N_2 \ , \quad
2l=N_2-N_3 \ , \quad
2r=N_3-n_0 \ . 
\label{6-6}
\eeq
Then, $\dket{lsr\omega}$ can be rewritten as  
\b\label{6-7}
\dket{lsr\omega}=({\hat q}^1)^{N_1-N_2}
({\hat S}^3)^{N_2-N_3}({\hat B}^*)^{N_3-n_0}({\hat b}^*)^{2(\Omega-n_0)}
\ket{0} \ . 
\end{equation}
In the expression (\ref{6-7}), the order of ${\hat q}^1$, ${\hat S}^3$ 
and ${\hat B}^*$ is changed. 
The eigenvalues of ${\hat H}$ and 
${\hat {\mib Q}}^2$ are expressed in 
terms of the new quantum numbers by substituting the expressions (\ref{6-5}) 
and (\ref{6-6}) to the relations (\ref{4-16b}) and (\ref{4-16c}).

\begin{figure}[b]
\begin{center}
\includegraphics[height=5.5cm]{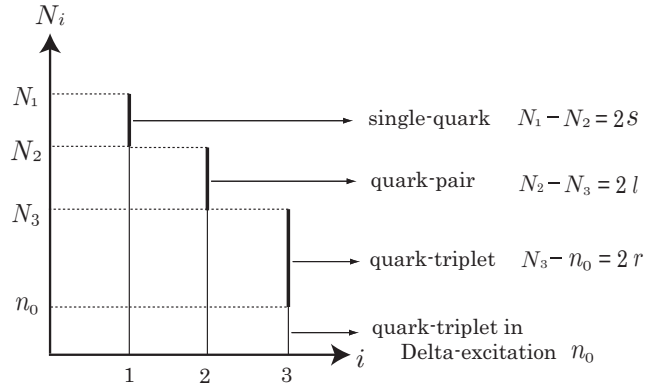}
\caption{The physical meaning of the inequality (\ref{5-9})
}
\label{fig:6-1}
\end{center}
\end{figure}

Since $2s$, $2l$, $2r$, $2\omega\geq 0$, we have the inequalities 
\beq
& &N_1 \geq N_2 \geq N_3 \geq n_0 \ , 
\label{6-8}\\
& &\Omega \geq n_0 \ . 
\label{6-9}
\eeq
Figure \ref{fig:6-1} shows the behaviors of $N_1$, $N_2$, $N_3$ 
and $n_0$. 
The quantities $(N_1-N_2)$, $(N_2-N_3)$, $(N_3-n_0)$ and $n_0$ denote 
the single-quark number in color 1, the quark-pair number 
in the pair (1,2), the quark-triplet number and the quark-triplet in 
$\Delta$-excitation in the sense of the idea proposed by Petry et al. 
in the state $\dket{lsr\omega}$. 
By varying $N_1$, $N_2$, $N_3$ and $n_0$, we obtain various phases. 
For example, in the case $N_1=N_2=N_3$ and $n_0=0$, the system 
consists of only the triplets. 
As the other extreme case, there exists the case $N_2=N_3=n_0=0$. 
In this case, the system consists only of the single-quarks. 
If $N_1=N_2$ and $N_3=n_0=0$, the system is composed only of the 
quark-pairs. 
Last two cases are in colored states. 
Of course, the first case is in color-neutral state. 
Anyhow, it may be interesting to investigate these various phases 
systematically.

Next, we investigate the eigenvalues of ${\hat \Omega}$, ${\hat N}_1$, 
${\hat N}_2$ and ${\hat N}_3$ calculated by the state (\ref{4-7}). 
The part $({\hat T}_+)^{\omega_0-\omega}$ does not give any effect 
because the sub-space with $\omega_0=\omega$, i.e., $T_0=T$ corresponds 
to the original fermion space. 
Judging 
from the role of the $su(1,1)$-algebra in the present formalism, 
it may be natural. 
Therefore, hereafter, we regard $\omega_0$ as $\omega$. 
The eigenvalue of ${\hat \Omega}$ does not change from $\Omega$ calculated 
in the state $\dket{lsr\omega}$. 
Therefore, the problem is reduced to investigating the effect 
of ${\hat Q}_+(lk\kappa\kappa_0)$ on $N_1$, $N_2$ and $N_3$. 
Straightforward calculation gives us 
\beq
& &[\ {\hat N}_1\ , \ {\hat Q}_+(lk\kappa\kappa_0)\ ]=
-2k{\hat Q}_+(lk\kappa\kappa_0) \ , 
\label{6-10}\\
& &[\ {\hat N}_2\ , \ {\hat Q}_+(lk\kappa\kappa_0)\ ]=
(k-(l+\kappa_0)){\hat Q}_+(lk\kappa\kappa_0) \ , 
\label{6-11}\\
& &[\ {\hat N}_3\ , \ {\hat Q}_+(lk\kappa\kappa_0)\ ]=
(k+(l+\kappa_0)){\hat Q}_+(lk\kappa\kappa_0) \ . 
\label{6-12}
\eeq
Therefore, the eigenvalues of ${\hat N}_1$, ${\hat N}_2$ and 
${\hat N}_3$ for the state ${\hat Q}_+(lk\kappa\kappa_0)\dket{lsr\omega}$, 
which are denoted as $N_1^0$, $N_2^0$ and $N_3^0$, respectively, 
are as follows: 
\beq
& &N_1^0=N_1-2k \ , 
\label{6-13}\\
& &N_2^0=N_2+k-(l+\kappa_0) \ , 
\label{6-14}\\
& &N_3^0=N_3+k+(l+\kappa_0) \ . 
\label{6-15}
\eeq
Of course, we have $\sum_i N_i^0=\sum_i N_i=N$. 
Energetically, the state 
${\hat Q}_+(lk\kappa\kappa_0)\dket{lsr\omega}$ is degenerate to 
$\dket{lsr\omega}$, but, the quark number distribution fluctuates 
around the distribution $(N_1, N_2, N_3)$ which determine the energy 
eigenvalue.

As a closing note in this section, we must mention a small comment. 
The state (\ref{4-1}) gives us the relation (\ref{6-8}). 
Then, we have the following question: 
How we treat the cases which do not obey the relation (\ref{4-1}), 
for example, $N_2 \ge N_1 \ge N_3$ ? 
If the colors $i=1$ and $i=2$ read the colors 
$i=2$ and $i=1$, respectively, we can treat this case in a way similar 
to the case (\ref{6-8}). 
The other cases are also same. 
Therefore, it may be enough to treat only the case (\ref{6-8}).

\section{Concluding remarks}

In this paper, we have reformulated the Bonn model and its modification 
developed in (I). 
We could show the energy eigenstates in a unified form and understood their 
structures in relation to the quark numbers in colors 
$i=1$, 2 and 3. 
In (I), we showed various inequalities for the physical quantities used 
in (I). 
Of course, these relations are available in the present form.

As a final remark, we will discuss a second modification of the 
original Bonn model. 
The Hamiltonian may be defined as 
\beq
{\hat H}_{m'}&=&{\hat H}+\chi{\hat {\mib Q}}^{2'} \ , 
\label{2-20}\\
{\hat {\mib Q}}^{2'}&=&
2({\hat S}_1^2{\hat S}_2^1+{\hat S}_1^3{\hat S}_3^1+{\hat R}_+{\hat R}_-)
\nonumber\\
&=&{\hat {\mib Q}}^2-\left[
\frac{2}{3}{\hat Q}_0({\hat Q}_0-3)+2{\hat R}_0({\hat R}_0-1)\right] \ , 
\label{2-21}
\eeq
where ${\hat {\mib Q}}^2$ is the Casimir operator of the $su(3)$ algebra and 
is defined in Eq.(\ref{2-13}). 
Then, we have 
\b\label{2-22}
[\ {\hat S}_i^j\ , \ {\hat H}_{m'}\ ] \neq 0 \ . \quad (i\neq j)
\end{equation}
Judging from the Hamiltonian (\ref{2-20}), we can diagonalize ${\hat H}_{m'}$ 
in the framework of the orthogonal set (\ref{4-7}) or (\ref{4-8}) with 
(\ref{4-1}). 
Of course, we are interested in the case $\omega_0=\omega$: 
$\dket{k\kappa\kappa_0 lsr\omega\omega_0=\omega}
=\dket{k\kappa\kappa_0 lsr\omega}$. 
We can calculate the energy eigenvalue of ${\hat {\mib Q}}^{2'}$ defined 
in the relation (\ref{2-21}) by using the relation (\ref{4-10a}). 
Then, the energy eigenvalue is expressed in the form 
\bsub\label{7-1}
\beq
& &E_{lsr\omega}^{(m')}=E_{lsr\omega}+\chi F_{ls}^{k\kappa\kappa_0} \ , 
\label{7-1a}\\
& &F_{ls}^{k\kappa\kappa_0}=2l(l+1)
+2k[(2l+4s+3)-3k]-2\kappa_0(\kappa_0-1)\ . 
\label{7-1b}
\eeq
\esub
Here, $E_{lsr\omega}$ is given in the form (\ref{4-10a}). 
The term $F_{ls}^{k\kappa\kappa_0}$ can be rewritten as 
\b\label{7-2}
F_{ls}^{k\kappa\kappa_0}=
4k(2s-k+1)+2(k+l-\kappa)(k+l+\kappa+1)
+2(\kappa+\kappa_0)(\kappa-\kappa_0+1) \ .
\end{equation}
The conditions (\ref{4-11b})$\sim$(\ref{4-11d}) lead us to 
$F_{ls}^{k\kappa\kappa_0}\geq 0$, which is, of course, consistent 
to the positive-definiteness of ${\hat {\mib Q}}^{2'}$ shown in the 
form (\ref{2-21}). 
The case $(k=0,\ \kappa=l,\ \kappa_0=-l)$ gives us 
$F_{ls}^{0l-l}=0$.

In the case of the Bonn model and its modification, the states 
$\dket{k\kappa\kappa_0 lsr\omega}$ are energetically degenerate to the state 
$\dket{lsr\omega}$ ($=\dket{0l-llsr\omega}$) and the energy 
eigenvalues are given in the form (\ref{4-16}). 
However, in the case the second modification, the degeneracy disappears. 
Only the state $\dket{lsr\omega}$ gives the energy eigenvalue $E_{lsr\omega}$ 
calculated in the frame of the Bonn model and the states 
$\dket{k\kappa\kappa_0 lsr\omega}$, which are orthogonal to 
$\dket{lsr\omega}$, become the excited states constructed on 
$\dket{lsr\omega}$. 
The above situation is very similar to the description of the 
rotational motion in axially symmetric deformed nuclei. 
In this case, the Hamiltonian is expressed in the form 
\beq\label{7-3}
{\hat H}&=&E({\hat I}_z^2)+\frac{1}{2{\cal G}}({\hat I}_x^2+{\hat I}_y^2) 
\nonumber\\
&=&E({\hat I}_z^2)+\frac{1}{2{\cal G}}{\hat {\mib I}}^2
-\frac{1}{2{\cal G}}{\hat I}_z^2 \ ,
\eeq
where $({\hat I}_x, {\hat I}_y, {\hat I}_z)$ denotes angular momentum 
operator and ${\hat {\mib I}}^2$ is the Casimir operator of 
the $su(2)$ algebra formed by this angular momentum operator. 
Here, $(x, y, z)$ shows the axes in the body-fixed frame. 
This operator obeys the relations $[{\hat I}_y, {\hat I}_z]=-i{\hat I}_x$ 
($x$, $y$, $z$: cyclic). 
The quantity ${\cal G}$ denotes the moment of inertia. 
The eigenvalue of ${\hat H}$ is given in the form 
\b\label{7-4}
E_{IK}=E(K^2)+\frac{1}{2{\cal G}}(I(I+1)-K^2) \ , \quad
I=K,\ K+1,\ K+2, \cdots \ (K\neq 0)
\end{equation}
If there does not exist the rotational term, the energy eigenvalue is 
given by $E(K^2)$ and all the states with $I=K,\ K+1,\ K+2,\cdots$ 
are degenerate. 
But, if the rotational term is switched on, the degenerate energies are 
splitted and the rotational states are constructed on the state with $I=K$. 
Usually, such sector forms the rotational band and the state with $I=K$ 
is called band head. 
In our present case, the state $\dket{lsr\omega}$ may be called as the band 
head and on this band head, the band structure is formed 
with the energy $F_{ls}^{k\kappa\kappa_0}$. 

As was shown in the above, the form discussed in this section may be 
permitted as a modified Bonn model. 
In the case of the nuclear rotational motion 
of the axially symmetric deformed nuclei, 
the ground state has no three-dimensional 
rotational symmetry which is described by the $su(2)$ algebra. 
However, the rotational symmetry around $z$ axis is rest and this symmetry 
is described by the $u(1)$ algebra. 
Actually, the Hamiltonian (\ref{7-3}) has still $u(1)$ symmetry, but 
no $su(2)$ symmetry. 
The same situation is realized in the modified Bonn model with the 
second modification represented by Eqs.(\ref{2-20}) and (\ref{2-21}). 
As is pointed out in \S 3, for the building blocks of the energy eigenstates, 
the operators ${\hat S}^3$ and ${\hat q}^1$ are not color $su(3)$ neutral. 
Therefore, the energy eigenstates are not the color $su(3)$ singlet 
in general. 
Further, the operator ${\hat S}^3$ is not $su(2)$ scalar, 
while ${\hat B}^*$ and ${\hat q}^1$ are still $su(2)$-scalars. 
Thus, the eigenstates has no longer the $su(2)$ symmetry as a sub-algebra 
of the color $su(3)$ algebra. 
However, the eigenstates are still 
possessed of a certain symmetry. 
Actually, the operators ${\hat Q}_0$ and ${\hat R}_0$ appearing in 
Eqs.(\ref{2-20}) and (\ref{2-21}) correspond to the two 
generators with only diagonal elements in the $su(3)$ algebra, namely 
form the Cartan-Weyl sub-algebra. 
Thus, the second modification in Eqs.(\ref{2-20}) and (\ref{2-21}) 
has only $u(1)\times u(1)$ symmetry while the eigenstates has no 
$su(3)$ and $su(2)$ symmetry. 
The above treatment may parallel the treatment of the nuclear rotational 
motion 
with axially symmetric deformed nuclei developed in the theory 
of nuclear collective motion.

\section*{Acknowledgements} 
One of the authors (Y.T.) 
is partially supported by the Grants-in-Aid of the Scientific Research 
No.18540278 from the Ministry of Education, Culture, Sports, Science and 
Technology in Japan.

\appendix

\section{Proofs of various formulae}
\subsection{Proof of the form (\ref{4-8}) with the conditions 
(\ref{4-11c}) and (\ref{4-11d})}

First, we note the following relations: 
\beq 
& &{\hat R}_-\dket{lsr\omega}=0 \ , \qquad
{\hat R}_0\dket{lsr\omega}=-l\dket{lsr\omega} \ , 
\label{a1}\\
& &{\hat S}_2^1\dket{lsr\omega}=0 \ , \qquad
{\hat S}_4^1\dket{lsr\omega}=0 \ . 
\label{a2}
\eeq
Here, ${\hat S}_4^1$ denotes Hermite conjugate of ${\hat S}_1^4$. 
Further, we have 
\beq
& &[\ {\hat R}_- \ , \ {\hat S}_1^2 \ ]=0 \ , \qquad
[\ {\hat R}_- \ , \ {\hat S}_1^4 \ ]={\hat S}_1^3{\hat R}_- \ , 
\label{a3}\\
& &[\ {\hat R}_0 \ , \ {\hat S}_1^2 \ ]=-\frac{1}{2}{\hat S}_1^2 \ , \qquad
[\ {\hat R}_0 \ , \ {\hat S}_1^4 \ ]=\frac{1}{2}{\hat S}_1^4 \ . 
\label{a4}
\eeq
The relations (\ref{a3}) and (\ref{a4}) give us 
\beq
& &[\ {\hat R}_- \ , \ ({\hat S}_1^2)^n \ ]=0 \ , \qquad
[\ {\hat R}_- \ , \ ({\hat S}_1^4)^n \ ]={\hat \Phi}_n\cdot{\hat R}_- \ , 
\label{a5}\\
& &[\ {\hat R}_0 \ , \ ({\hat S}_1^2)^n \ ]
=-\frac{n}{2}({\hat S}_1^2)^n \ , \qquad
[\ {\hat R}_0 \ , \ ({\hat S}_1^4)^n \ ]=\frac{n}{2}({\hat S}_1^4)^n \ . 
\label{a6}
\eeq
The operator ${\hat \Phi}_n$ denotes a certain function of 
${\hat S}_1^3$ and ${\hat S}_1^4$, but, in the present argument, 
the explicit form is not necessary.

With the use of the above relations, we obtain the relation 
\beq
& &{\hat R}_-\dket{k\kappa lsr\omega}=0 \ , \qquad
{\hat R}_0\dket{k\kappa lsr\omega}=-\kappa\dket{k\kappa lsr\omega} \ , 
\label{a7}\\
& &\kappa \geq 0 \ . 
\label{a8}
\eeq
Here, $\dket{k\kappa lsr\omega}$ is defined as 
\b\label{a9}
\dket{k\kappa lsr\omega}=({\hat S}_1^2)^{k-l+\kappa}({\hat S}_1^4)^{k+l-\kappa}
\dket{lsr\omega} \ . 
\end{equation}
The state $\dket{k\kappa lsr\omega}$ is nothing but the minimum weight state 
of $({\hat R}_{\pm,0})$, in which $-\kappa$ denotes the eigenvalue of 
${\hat R}_0$. 
This state is also the minimum weight state of $({\hat T}_{\pm,0})$, 
in which $\omega$ denotes the eigenvalue of ${\hat T}_0$. 
Then, with the use of the raising operators 
${\hat R}_+$ and ${\hat T}_+$, we have the form (\ref{4-8}) with the condition 
(\ref{4-11d}).

We, further, note the relation 
\beq\label{a10}
& &{\hat R}_-({\hat S}_1^4)^{k+l-\kappa} \dket{lsr\omega}=0 \ , \nonumber\\
& &{\hat R}_0({\hat S}_1^4)^{k+l-\kappa} \dket{lsr\omega}
=-\frac{1}{2}(l-k+\kappa)({\hat S}_1^4)^{k+l-\kappa} \dket{lsr\omega} \ . 
\eeq
We can see that $({\hat S}_1^4)^{k+l-\kappa}\dket{lsr\omega}$ is also the 
minimum weight state of $({\hat R}_{\pm,0})$ with the eigenvalue 
of ${\hat R}_0$ 
being $-(l-k+\kappa)/2$. 
Therefore, in the same way as the condition (\ref{a8}), we have 
\b\label{a11}
l-k+\kappa \geq 0 \ . 
\end{equation}
The exponents $(k-l+\kappa)$ and $(k+l-\kappa)$ in the state (\ref{a9}) 
should be positive or zero: 
\b\label{a12}
k-l+\kappa \geq 0 \ , \qquad k+l-\kappa \geq 0 \ . 
\end{equation}
The inequalities (\ref{a11}) and (\ref{a12}) lead us to the 
condition (\ref{4-11c}). 
As is clear from the relation (\ref{4-10a}), $k$ is related 
to the eigenvalue of ${\hat Q}_0$. 
Since $[{\hat S}_1^2, {\hat S}_1^4]=0$, we have 
\b\label{a13}
\dket{k\kappa lsr\omega}=({\hat S}_1^4)^{k+l-\kappa}
({\hat S}_1^2)^{k-l+\kappa}\dket{lsr\omega} \ .
\end{equation}
The case $({\hat S}_1^2)^{k-l+\kappa}\dket{lsr\omega}$ is also 
treated in the same form as the case 
$({\hat S}_1^4)^{k+l-\kappa}\dket{lsr\omega}$. 
In this case, the eigenvalue of ${\hat R}_0$ is given as $-(k+l+\kappa)/2$, 
which is automatically negative. 
Therefore, any condition is not obtained. 
The idea presented in this Appendix is very similar to that 
developed by the present authors.

\subsection{Derivation of the eigenvalue of ${\hat {\mib P}}^2$ shown 
in the relation (\ref{4-10})}

First, we notice that straightforward calculation gives us the following expression: 
\beq\label{a14}
\dket{k\kappa\kappa_0lsr\omega\omega_0}
&=&({\hat T}_+)^{\omega_0-\omega}({\hat R}_+)^{\kappa+\kappa_0}
({\hat S}_1^2)^{k-l+\kappa}({\hat S}_1^4)^{k+l-\kappa} 
\nonumber\\
& &\times ({\hat S}^3)^{2l}{\hat P}_+(sr\omega)\dket{sr\omega} \ .
\eeq
Here, ${\hat P}_+(sr\omega)$ and $\dket{sr\omega}$ are defined as 
\beq 
{\hat P}_+(sr\omega)&=&
\left(\begin{array}{@{\,}c@{\,}}
2\omega \\
2(s+r)
\end{array}\right)
\sum_{(n)}{}'\frac{(2r)!(2s+n_1)!}{n_1!n_2!n_3!} \nonumber\\
& &\times ({\hat S}^1)^{n_1}({\hat S}^2)^{n_2}({\hat S}_1^2)^{n_2}
({\hat S}^3)^{n_3}({\hat S}_1^3)^{n_3} \ , 
\label{a15}\\
\dket{sr\omega}&=&({\hat q}^1)^{2(s+r)}({\hat b}^*)^{2\omega}\ket{0} \ .
\label{a16}
\eeq
The symbol $\sum_{(n)}{}'$ denotes the sum restricted 
$n_1+n_2+n_3=2r$. 
If we note that the $su(1,1)$- and the $su(4)$-generators 
commute with ${\hat {\mib P}}^2$ and the state $\dket{sr\omega}$ is 
the minimum weight state for the $su(4)$-algebra, 
the eigenvalue of ${\hat {\mib P}}^2$ for 
$\dket{k\kappa\kappa_0 sr\omega\omega_0}$ is equal to that for 
$\dket{sr\omega}$ and we have 
\b\label{a17}
{\hat {\mib P}}^2\dket{sr\omega}=
[(2(s+r)-\omega)^2+2\omega(\omega+3)]\dket{sr\omega} \ .
\end{equation}
The above result is nothing but the result shown in the relation (\ref{4-10}).

\subsection{Derivation of the inequality (\ref{4-11a})}

The state $\dket{lsr\omega}$ can be rewritten as 
\b\label{a18}
\dket{lsr\omega}=({\hat B}^*)^{2r}({\hat a}_3^*)^{2l}
({\hat b}_1^*)^{2s}({\hat b}^*)^{2(\omega-(l+s))}\ket{0} \ .
\end{equation}
The operator ${\hat B}^*$ can be expressed explicitly as follows: 
\b\label{a19}
{\hat B}^*={\hat \tau}_+{\hat b}^2-2{\hat \tau}_0{\hat a}^*{\hat b}
+{\hat \tau}_-{\hat a}^{*2} \ .
\end{equation}
Noticing ${\hat \tau}_-({\hat a}_3^*)^{2l}({\hat b}_1^*)^{2s}
({\hat b}^*)^{2(\omega-(l+s))}\ket{0}=0$, the expression 
(\ref{a18}) can be rewritten in the form 
\beq\label{a20}
\dket{lsr\omega}&=&[{\hat c}_0\cdot({\hat b})^{4r}
+{\hat c}_1\cdot{\hat a}^*({\hat b})^{4r-1}+\cdots +
{\hat c}_{2r}\cdot ({\hat a}^*)^{2r}({\hat b})^{2r}] \nonumber\\
& &\times 
({\hat a}_3^*)^{2l}({\hat b}_1^*)^{2s}({\hat b}^*)^{2(\omega-(l+s))}\ket{0} \ .
\eeq
Here, ${\hat c}_k$ ($k=0,\ 1,\cdots ,\ 2r)$ denotes a certain function 
of ${\hat \tau}_+$ and ${\hat \tau}_0$. 
In the present argument, the concrete expression is not necessary to show. 
Then, the relation (\ref{a20}) shows us that the state $\dket{lsr\omega}$ 
does not vanish if there exists the condition 
\b\label{a21}
2(\omega-(l+s))-2r \geq 0 \ , \quad
{\rm i.e.,}\quad 
l+s+r \leq \omega\ .
\end{equation}
The relation (\ref{a21}) is nothing but the inequality (\ref{4-11a}). 
We can see that under the condition (\ref{a21}), the exponent 
$2(\omega-(l+s))$ in the relation (\ref{a18}) is positive 
or zero.

\subsection{Derivation of the inequality (\ref{4-11b})}

In (I), the following inequality was presented in the relation 
(I$\cdot$3$\cdot$17): 
\b\label{a22}
2(l+k) \leq 2\tau-3 \ .
\end{equation}
The notations are changed from the original. 
The proof was given in Appendix B of (I). 
On the other hand, in the relation (\ref{5-22}), $(2\tau-3)$ is shown 
in the form
\b\label{a23}
2\tau-3=2(s+l) \ .
\end{equation}
Then, we have 
\b\label{a24}
2(l+k) \leq 2(s+l) \ , \quad {\rm i.e.,}\quad 
k\leq s \ .
\end{equation}
The relation (\ref{a24}) is nothing but the inequality (\ref{4-11b}).

\end{document}